\documentclass[apjl]{emulateapj}
\usepackage{apjfonts}
\usepackage{times,color}

\shorttitle{Turbulence and Magnetic Field Amplification in Supernova Remnants}
\shortauthors{Inoue et al.}

\begin{document}

\title{
Turbulence and Magnetic Field Amplification in Supernova Remnants: Interactions Between A Strong Shock Wave and Multi-Phase Interstellar Medium
}
\author{Tsuyoshi Inoue\altaffilmark{1}, Ryo Yamazaki\altaffilmark{2}, and Shu-ichiro Inutsuka\altaffilmark{3}}
\altaffiltext{1}{Division of Theoretical Astronomy, National Astronomical Observatory of Japan, Osawa, Mitaka 181-8588 Japan; inouety@th.nao.ac.jp}
\altaffiltext{2}{Department of Physical Science, Hiroshima University, Higashi-Hiroshima, Hiroshima 739-8526, Japan}
\altaffiltext{3}{Department of Physics, Graduate School of Science, Kyoto University, Sakyo-ku, Kyoto 606-8588, Japan}

\begin{abstract}
We examine MHD simulations of the propagation of a strong shock wave through the interstellar two-phase medium composed of small-scale cloudlets and diffuse warm neutral medium in two-dimensional geometry.
The pre-shock two-phase medium is provided as a natural consequence of the thermal instability that is expected to be ubiquitous in the interstellar medium.
We show that the shock-compressed shell becomes turbulent owing to the preshock density inhomogeneity and magnetic field amplification takes place in the shell.
The maximum field strength is determined by the condition that plasma $\beta\sim 1$, which gives the field strength on the order of 1 mG in the case of shock velocity $\sim 10^3$ km s$^{-1}$.
The strongly magnetized region shows filamentary and knot-like structures in two-dimensional simulations.
The spatial scale of the regions with magnetic field of $\sim$1 mG in our simulation is roughly 0.05 pc which is comparable to the spatial scale of the X-ray hot spots recently discovered in supernova remnants where the magnetic field strength is indicated to be amplified up to the order of 1 mG.
This result may also suggest that the turbulent region with locally strong magnetic field is expected to be spread out in the region with frequent supernova explosions, such as in the Galactic center and starburst galaxies. 
\end{abstract}


\section{Introduction}
Recent discovery of the year-scale variability in the synchrotron X-ray emission of supernova remnants (SNRs) suggests that the magnetic field should be amplified in the SNR up to the level of milli-Gauss (Uchiyama et al. 2007; Uchiyama \& Aharonian 2008, see also Bamba et al. 2003, 2005a, 2005b; Vink \& Laming 2003).
Since the typical magnetic field strength in the interstellar medium (ISM) is on the order of micro-Gauss, amplification beyond the simple shock compression is necessary to achieve a milli-Gauss level of magnetic field in a supernova shell.

The amplification of magnetic field around the shock wave has been one of the main interests of plasma physics.
Recently Giacalone \& Jokipii (2007) have demonstrated by using magnetohydrodynamic (MHD) simulation that density fluctuations in the preshock medium cause turbulence and magnetic field amplification in the postshock medium.
The density fluctuations in their preshock state are given by a lognormal probability distribution with a Kolmogorov-like power spectrum, and the maximum magnetic field strength achieved in the postshock medium is larger than a hundred times the preshock field strength.
The density fluctuations with Kolmogorov power spectrum would be accomplished in the preshock medium owing to the transonic nature of turbulence expected in the diffuse ISM (e.g., Hennebelle \& Audit 2007).
However, applicability of the lognormal distribution of density fluctuations for the ISM is unknown, although their pioneering study is significant. 
Balsara et al. (2001) have studied the case of expanding supernova blast wave in the turbulent ISM, but have not found the magnetic field amplification beyond 50 $\mu$G.
Thus, the turbulent field alone does not seem to be able to amplify magnetic field to the level of milli-Gauss.

It has been known that the ISM is not a uniform but a highly inhomogeneous medium composed of clumpy HI clouds embedded in diffuse intercloud medium.
Observations through the 21cm absorption lines have shown that HI clouds with column densities $N\sim 10^{19}$ cm$^{-2}$, whose spatial scale corresponds to $l\sim 0.1$ pc, are ubiquitous in the ISM (see, e.g., Heiles \& Troland 2003).
The coexistence of the HI clouds and intercloud gas can be understood as follows:
In the ISM, the balance between the line-emission coolings and the heating due to ultraviolet background radiation determines two thermally stable equilibrium states with different temperatures that can coexist in pressure balance (Field et al. 1969; Wolfire et al. 1995).
One of the equilibrium states corresponds to the diffuse intercloud medium, the so-called warm neutral medium (WNM; $n\sim 0.5$ cm$^{-3}$, $T\sim 8,000$ K), and the other one corresponds to the HI clouds, the so-called cold neutral medium (CNM; $n\sim 50$ cm$^{-3}$, $T\sim 100$ K).
Thus, the coexistence of HI clouds and surrounding diffuse intercloud gas is naturally achieved by the thermally bistable nature of the ISM.
The high-resolution MHD simulation performed by Inoue \& Inutsuka (2008) has shown that when a weak shock wave ($v_{\rm shock}\sim 20$ km s$^{-1}$) propagates into the diffuse WNM, it generates thermally unstable gas that evolves into the small scale cloudlets of CNM embedded in the WNM via the thermal instability (Field 1965).
The scale of the CNM cloudlets generated by the thermal instability is typically 0.1 pc that resembles the HI clouds observed in 21cm absorption lines.
Since the ISM is frequently swept up by weak shock waves, e.g., due to old supernova blast waves at a rate on the order of once per millions of years (McKee \& Ostriker 1977), the typical ISM should be treated as a ``two-phase medium" composed of the WNM and CNM rather than a single phase polytropic gas.

In this paper, we examine the propagation of strong shock waves into the two-phase medium by using two-dimensional MHD simulation, in which the shock strength corresponds to the Sedov-Taylor phase of the SNR. 
In \S 2 we provide basic assumptions and numerical methods to generate the two-phase medium and to induce shock waves.
The results of the simulations are shown in \S 3.
Finally, in \S 4, we summarize our results and discuss the implications.

\section{Basic Assumptions and Numerical Methods}
We solve the MHD equations for Cartesian geometry in a conservative fashion:
\begin{eqnarray}
&&\frac{\partial\rho}{\partial t}+\vec{\nabla}\cdot(\rho\,\vec{v})=0,\\
&&\frac{\partial\rho\,\vec{v}}{\partial t}+\vec{\nabla}\cdot(p+\frac{B^2}{8\,\pi}+\rho\,\vec{v}\otimes\vec{v}-\frac{1}{4\,\pi}\,\vec{B}\otimes\vec{B})=0,\\
&& \frac{\partial e}{\partial t}+\vec{\nabla}\cdot\{(e+p+\frac{B^2}{8\,\pi})\,\vec{v}-\frac{\vec{B}\cdot\vec{v}}{4\,\pi}\vec{B}\}=\vec{\nabla}\cdot\kappa\vec{\nabla} T-L(n,T),\\
&&\frac{\partial \vec{B}}{\partial t}=\vec{\nabla}\times(\vec{v}\times\vec{B}),\\
&&e=\frac{p}{\gamma-1}+\frac{\rho\,v^2}{2}+\frac{B^2}{8\,\pi},
\end{eqnarray}
where $\kappa$ is the thermal conductivity and $L(n,T)$ is the net cooling function whose details are noted in the following section.
We impose the ideal gas equation of state to close the equations.
We use the second-order Godunov scheme (van Leer 1979) for solving MHD equations, in which the hydrodynamic equations with the magnetic pressure term are solved based on the solution of the Riemann problem (Sano et al. 1999), the magnetic tension terms are solved using the method of characteristics for Alfv\'en waves (MOC: Stone \& Norman 1992), and the induction equations are solved using the consistent MOC with the constrained transport algorithm (Clark 1996).
We use the second-order explicit time integration for the cooling/heating and thermal conduction.
The two-dimensional computational domain of $x\times y=2^2$ pc$^2$ ($0\le x,y \le 2$ pc) is used with the uniform $4096^2$ cells ($\Delta x=\Delta y=4.9\times 10^{-4}$ pc).

Computations are performed in the following two stages:
(1) the generation stage of the two-phase medium by the thermal instability, and (2) the propagation stage  of the shock wave through the two-phase medium.
In the following we describe the detailed numerical settings of each stage.

\subsection{Stage 1: Generation of Pre-Shock Two-Phase Medium}
\begin{figure}
\epsscale{1.}
\plotone{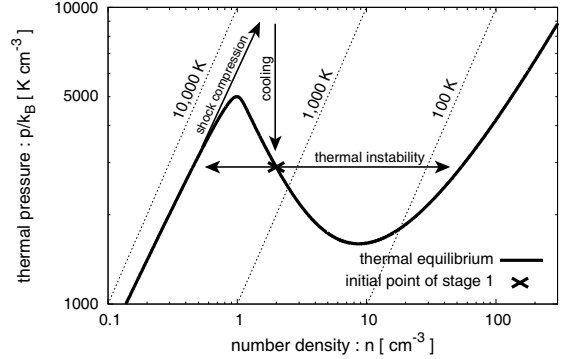}
\caption{
Thermal equilibrium state of the cooling function ({\it thick solid line}).
The isothermal lines of 10,000 K, 1,000 K, and 100 K are plotted as dashed lines.
The evolutionary track of diffuse WNM that compressed by a shock wave and cooled by following radiative cooling to form thermally unstable gas is illustrated as allows.
We also illustrate the evolutionary tracks of the thermal instability that are simulated in stage 1.
The initial unperturbed state of stage 1 is plotted as a cross.
}
\label{f1}
\end{figure}

In stage 1, we take into account the effects of the cooling/heating and thermal conduction.
We use the cooling/heating functions given by Koyama \& Inutsuka (2002) that are obtained by fitting various line-emission coolings (CII $158 \mu$m, OI $63 \mu$m, etc.) and photoelectric heating from dusts, which can adequately describe the effects of cooling/heating in the ISM in the range $10$ K $\lesssim T \lesssim 10^4$  K. 
The thermal equilibrium state with this cooling/heating is shown in Fig. \ref{f1}.
Since this stage deals with weakly ionized medium, the isotropic thermal conductivity due to the neutral atomic collisions ($\kappa = 2.5\times 10^3$ erg cm$^{-1}$ s$^{-1}$ K$^{-1}$: Parker 1953) is used.

In order to generate the two-phase medium, we initially prepare a uniform gas in thermally unstable equilibrium with $n=2.0$ cm$^{-3}$ and $p/k_{\rm B}=2900$ K cm$^{-3}$ (the initial point is plotted in Fig. \ref{f1} as a cross).
Flat-spectrum density perturbations are added to the unstable gas to seed the thermal instability.
The initial magnetic field strength is uniform with the orientation parallel to the $x$-axis and the strength of $B_0=6.0$ $\mu$G.
Note that, as shown in Inoue \& Inutsuka (2008), such a condition is naturally produced by the shock compression of the diffuse WNM and the subsequent isochoric radiative cooling.
In Fig. \ref{f1} we illustrate a schematic evolutionary track from the WNM to thermally unstable medium based on the results of Inoue \& Inutsuka (2008).
We use the periodic boundary conditions.
The calculation of the first stage is stopped at $t=4.0$ Myr.
Since the growth timescale of the thermal instability is approximately equal to the cooling timescale that is approximately a few Myr in the typical ISM, a typical two-phase medium is formed at $t=4$ Myr as a consequence of the thermal instability.

\subsection{Stage 2: Injection of Shock Wave}
In stage 2, we examine propagation of a shock wave through the two-phase medium generated as a result of stage 1.
We consider the propagation of parallel shock and perpendicular shock.
In the case of the parallel shock, we set the hot plasma with $n=0.1$ cm$^{-3}$ at the boundary $x=0.0$ by which the shock wave is induced.
The free boundary condition is imposed at $x=2.0$ pc, and we impose the periodic boundary conditions for the boundaries at $y=0.0$ and $2.0$ pc.
In the case of the perpendicular shock, we set the hot plasma at the boundary $y=0.0$.
The free boundary condition is imposed at $y=2.0$ pc, and the periodic boundary conditions are imposed for the boundaries at $x=0.0$ and $2.0$ pc.
We measure the time since the shock is induced.

We study the effect of the shock strength by changing thermal pressure of the hot plasma from $p_{\rm th}/k_{\rm B}=10^{8}$ to $10^{7}$ K cm$^{-3}$.
These models are summarized in Table \ref{t1}.
For convenience, we list the resulting average propagation speeds of the shocks $v_{\rm shock}$ in the fourth column of Table \ref{t1}.
The computations are stopped when the shock front reaches the opposite boundary.
The effects of the thermal conduction, cooling, and heating that considered in stage 1 are omitted in this stage, since the dynamical timescale of this stage ($\sim 1,000$ yr) is much shorter than the timescales of these effects ($\sim 1$ Myr).
The hydrodynamic treatment would be reasonable on the scale considered in this paper, since the gyration radius for a thermal proton is estimated as
\begin{eqnarray}
l_{\rm g}&=&1.5\times 10^9\,\left( \frac{p/k_{\rm B}}{3\times 10^8\,\,\rm{K cm}^{-3}} \right)^{1/2}\,
\left( \frac{n}{10\,\,\rm{ cm}^{-3}} \right)^{-1/2}\,\nonumber\\
&&\times\left( \frac{B}{6\,\,\mu\rm{G}} \right)^{-1}\,\,\rm{cm},
\end{eqnarray}
which is much smaller than the spatial resolution of our numerical simulations and justifies (magneto-)hydrodynamics approximation. 

\begin{deluxetable}{llll}
\tablewidth{0pt}
\tablecaption{Model Parameters}
\tablehead{Model & $p_{\rm th}/k_{\rm B}$& shock type & shock speed}
\startdata
1..... & $3.0\times 10^{8}\,$K cm$^{-3}$ & perpendicular shock & 1256 km s$^{-1}$ \\
2..... & $3.0\times 10^{8}\,$K cm$^{-3}$ & parallel shock & 1289 km s$^{-1}$\\
3..... & $1.0\times 10^{8}\,$K cm$^{-3}$ & parallel shock & 726 km s$^{-1}$\\
4..... & $3.0\times 10^{7}\,$K cm$^{-3}$ & parallel shock & 397 km s$^{-1}$\\
\enddata
\label{t1}
\end{deluxetable}

\section{Results and Interpretations}
\subsection{Results of Stage 1: Two-Phase Medium}
\begin{figure}[h]
\epsscale{1.3}
\plotone{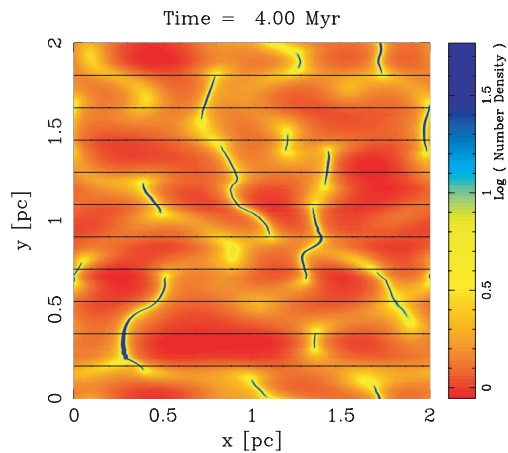}
\caption{
Resulting density structure of the stage 1 at $t=4.0$ Myr.
The black lines show magnetic field lines.
}
\label{f2}
\end{figure}

The resulting density structure of the stage 1 is shown in Fig. \ref{f2}.
The cold and dense filamentary clumps (CNM) and the surrounding warm diffuse gas (WNM) are formed as a consequence of the thermal instability.
The condensation driven by the thermal instability to form the CNM essentially arises along the magnetic field, since the motion perpendicular to the field line is easily stopped owing to the enhancement of the magnetic pressure.
Thus, the thickness of the filaments can be described by the most unstable scale of the thermal instability ($\sim 1$ pc) times the compression ratio of the condensation that gives $\sim 0.1$ pc, and the length of the filament is roughly given by the most unstable scale of the thermal instability.

We stress that such a two-phase structure is quite natural and ubiquitously expected in the ISM not only from the theoretical but also from the observational point of view (Heiles \& Troland 2003).
A more detailed description of the evolution of the thermal instability can be found, e.g., in Inoue et al. (2007), Hennebelle \& Audit (2007), and Inoue \& Inutsuka (2008).

\subsection{Results of Stage 2: Effects of Shock Wave}
\subsubsection{Turbulence in The Shell}\label{s31}
\begin{figure*}
\epsscale{1.1}
\plotone{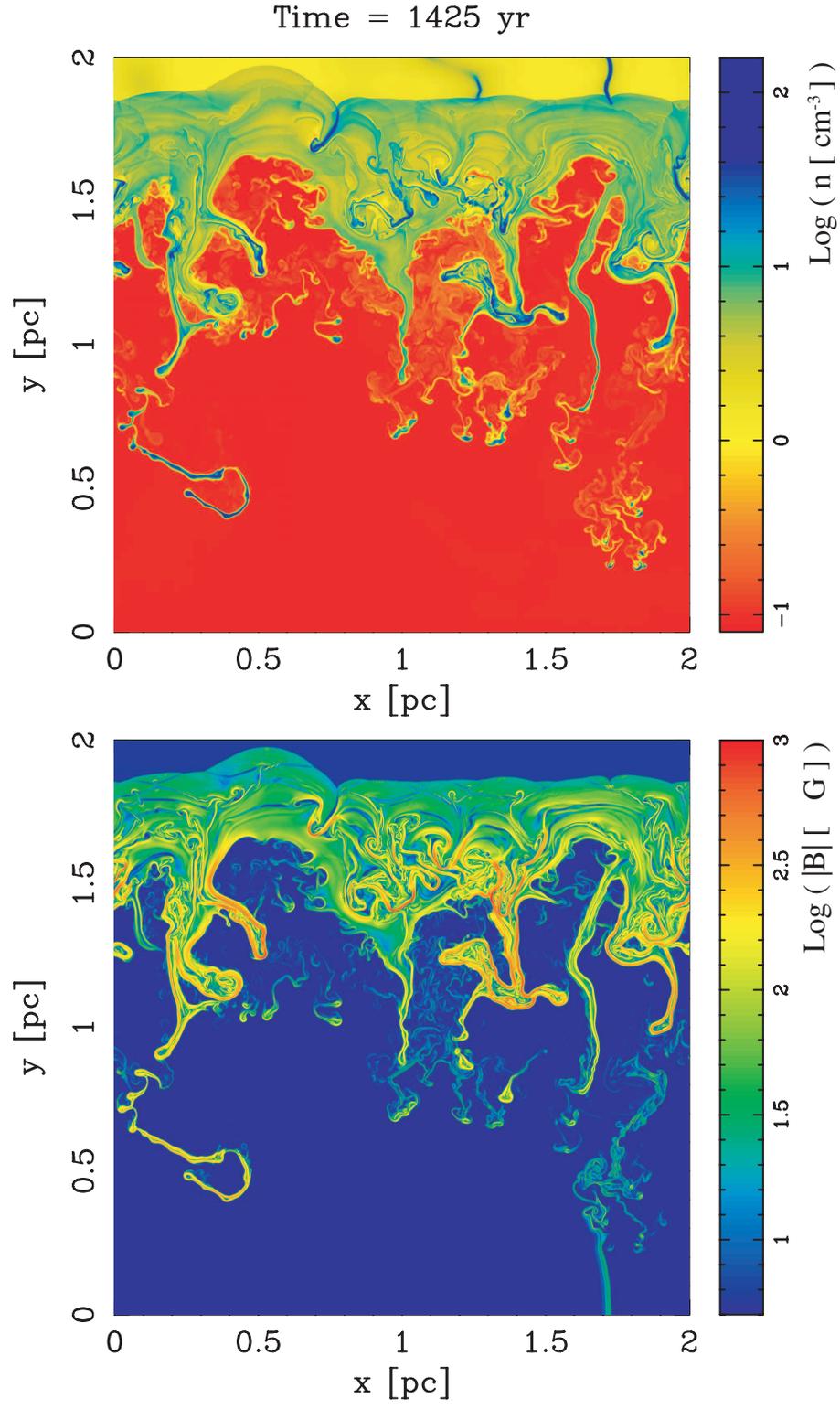}
\caption{
Results of Model 1.
The top and bottom panels respectively represent the structure of the number density and magnetic field strength.
}
\label{f3}
\end{figure*}

\begin{figure*}
\epsscale{1.1}
\plotone{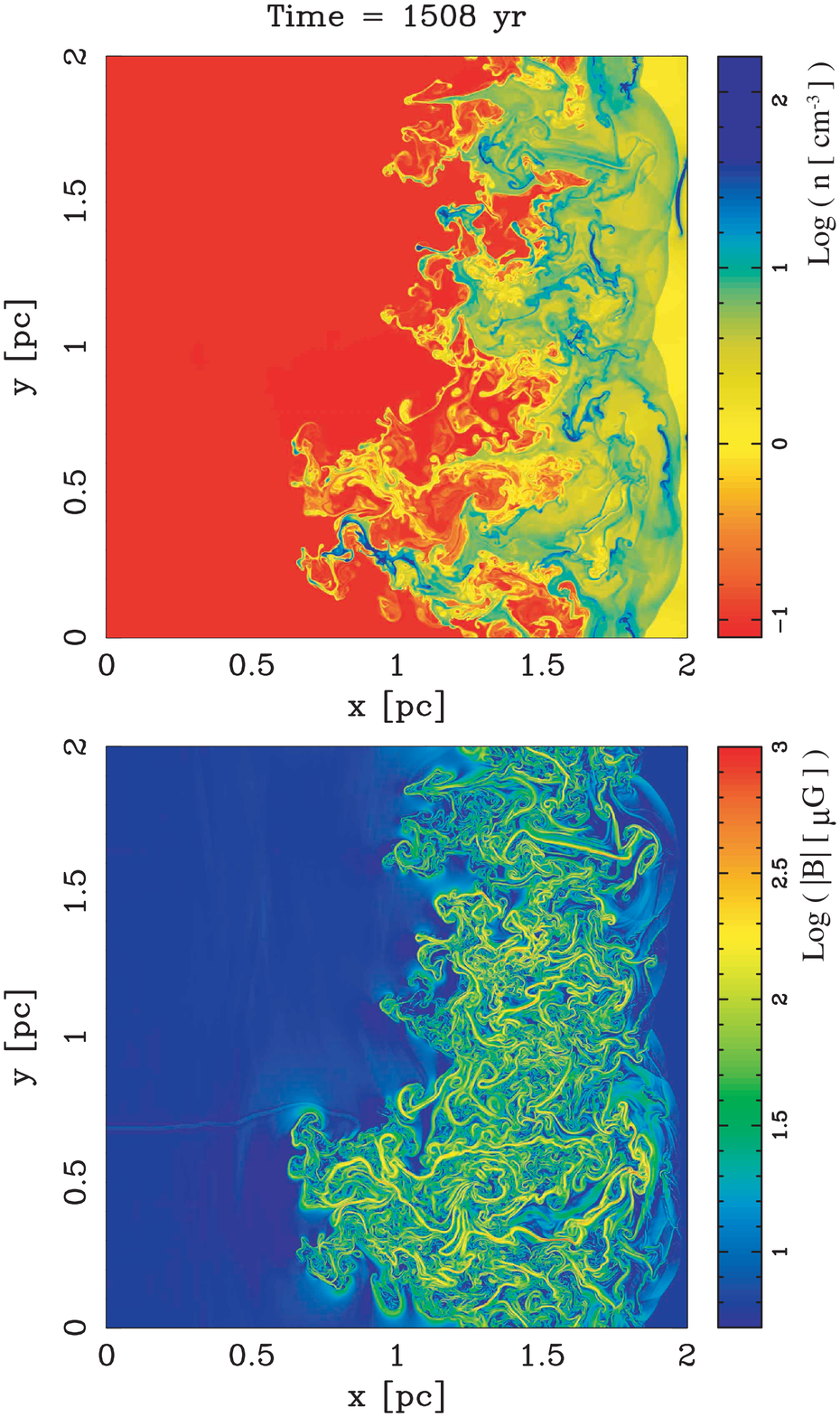}
\caption{
Results of Model 2.
The top and bottom panels respectively represent the structure of the number density and magnetic field strength.
The color scale is the same as Fig. \ref{f3}.
}
\label{f4}
\end{figure*}

In Fig. \ref{f3} and \ref{f4} we show the results of Model 1 at $t=1,425$ yr and Model 2 at $t=1,508$ yr, respectively.
The top and bottom panels respectively represent the structure of the number density and magnetic field strength. 
In the following we use data at these snapshots for the analyses of Models 1 and 2.

In both Models 1 and 2, the Alfv\'en Mach number $M_{\rm A}$ and the sonic Mach number $M_{\rm s}$ in the WNM and CNM can be roughly estimated as
\begin{eqnarray}
M_{\rm A, WNM}&\sim& \frac{v_{\rm shock}}{B_0/\sqrt{4\,\pi\,\rho_{\rm WNM}}}\simeq 110,\\
M_{\rm s, WNM}&\sim& v_{\rm shock}/c_{\rm s, WNM}\simeq 170,\\
M_{\rm A, CNM}&\sim& \frac{v_{\rm shock}}{B_0/\sqrt{4\,\pi\,\rho_{\rm CNM}}}\simeq 760,\\
M_{\rm s, CNM}&\sim& v_{\rm shock}/c_{\rm s, CNM}\simeq 1200,
\end{eqnarray}
where we have used the conditions $n_{\rm CNM}=50$ cm$^{-3}$, $T_{\rm CNM}=100$ K, $n_{\rm WNM}=1$ cm$^{-3}$, and $T_{\rm CNM}=5000$ K in the evaluation.

The density structures in both models indicate that the shocked shells that are formed by piling up the two-phase medium are turbulent.
As discussed in Giacalone \& Jokipii (2007), the density bumps in the preshock region lead to the rippling of the shock front that consequently generates vortexes in the post shock flow. 
This type of vorticity generation is known as Richtmyer-Meshkov instability that is a kind of Rayleigh-Taylor type instability (e.g., Brouillette 2002).
Many mushroom-shaped structures that indicate the nonlinear growth of the Rayleigh-Taylor type instability are found in the shell (see, top panels of Fig. \ref{f3} and Fig. \ref{f4}).

In order to adequately calculate the rippling of the shock front that determines the vorticity of the post shock flow, we have to resolve the transition layers between the WNM and the CNM clumps at which the curvature of the rippling is the largest and the generated vorticity is the strongest.
It is known that the scale of the transition layer is determined by so called ``Field length'' $l_{\rm F}=\sqrt{\kappa\,T/L(n,T)}$ where $L(n,T)$ is the cooling rate per unit volume (Begelman \& McKee 1990).
In the typical ISM, the Field length takes the value of $l_{\rm F}\sim 0.01-0.1$ pc (Inoue et al. 2006), which is much larger than the resolution of our simulations.
Thus, our simulations can describe the post shock turbulence accurately.

The velocity dispersions in the shocked shells ($\delta v = 500$ km s$^{-1}$ for Model 1 and $450$ km s$^{-1}$ for Model 2) are comparable to the sound speed in the shell:
\begin{equation}\label{ss}
c_{\rm s} = 570\,\left( \frac{p/k_{\rm B}}{ 3\times10^{8}\,{\rm K cm}^{-3}} \right)^{1/2}\left( \frac{n}{10\,{\rm cm}^{-3}} \right)^{-1/2}\,{\rm km\,\,s}^{-1},
\end{equation}
where the velocity dispersions are estimated by the half widths of the velocity distribution in the shell.
Throughout this paper, we define the region in ``the shell'' such that $p/k_{\rm B}>10^{5}$ K cm$^{-3}$ and $n>1$ cm$^{-3}$ in the region. 
The former condition excludes the preshock region and the latter condition excludes the region filled by hot plasma with $n\simeq 0.1$ cm$^{-3}$.

\subsubsection{Magnetic Field Amplification}\label{s32}
The magnetic field strength in the shell has huge amplitude fluctuations as depicted in the bottom panels of Fig. \ref {f3} and Fig. \ref{f4}.
There exist many filamentary regions with the field strength far beyond the value expected only from the shock compression.
The maximum field strengths achieved during the evolutions are $1280\,\mu$G in Model 1 and $860\,\mu$G in Model 2.

\begin{figure}[t]
\epsscale{1.}
\plotone{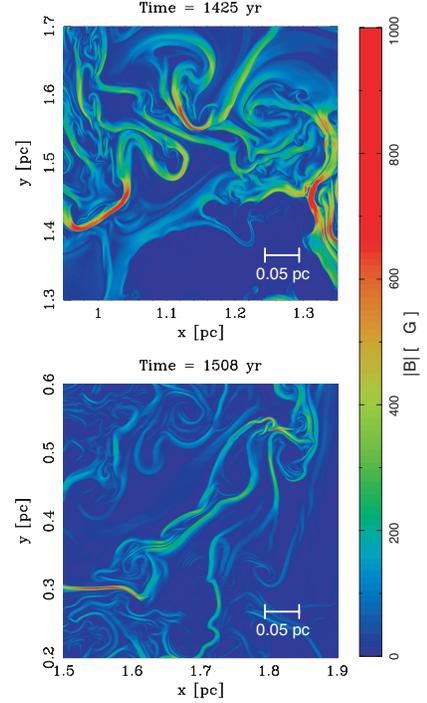}
\caption{
Close-up views of the magnetic field strength distribution of Model 1 ({\it top}) and 2 ({\it bottom}).
The regions depicted in red have field strength larger than 600 $\mu$G.
}
\label{f5}
\end{figure}

The amplifications of the magnetic fields would come from the turbulence.
Since the preexisting magnetic field is much weaker than the post shock turbulence, the turbulent velocity field can easily stretch and deform the magnetic field lines, which creates the regions with the amplified magnetic field (Giacalone \& Jokipii 2007).
One can further understand the mechanism of the strong field amplification by the following equation that is obtained from the equation of continuity and the induction equation:
\begin{equation}
\frac{d}{dt}\left( \frac{\vec{B}}{\rho} \right)=\frac{1}{\rho}\,(\vec{B}\cdot\vec{\nabla})\,\vec{v}.
\end{equation}
This equation shows that the magnetic field can be amplified, if the velocity has shear along the field line.
Such a situation is realized in the post shock shell due to the turbulence.
Especially at the transition layer between CNM and WNM, the velocity shear (vortex) is induced most strongly, since the propagation velocity of the shock wave is very different  in WNM and CNM. 
This picture of the magnetic field amplification is reinforced by the following properties of the shocked shell.
First, the regions where the magnetic field is strongly amplified roughly traces the transition layers between shocked CNM clumps and WNM.
Fig. \ref{f5} shows the closeup views of the magnetic field strength structure of Model 1 ({\it top}) and 2 ({\it bottom}).
One can see that the thickness of  the regions with strong magnetic field is approximately 0.01 pc, which agrees well with the thickness of the transition layer (Inoue et al. 2006).
Second, in the region where the strength of the magnetic field is maximum, the plasma $\beta$ is on the order of unity (see, \S \ref{ps} below).
This reflects the fact that the back reaction of the magnetic field stops amplification due to turbulent flows whose velocity dispersion is comparable to the sound speed.

In the top panel of Figure \ref{f9}, we plot time evolutions of maximum ({\it solid}) and average ({\it dotted}) field strength in the shocked shell, where the average field strength is defined by $\langle |\vec{B}|  \rangle_{\rm shell}$.
The red and green lines respectively represent those of Model 1 (perpendicular shock) and 2 (parallel shock). 
The maximum field strengths grow fast and saturate at $\sim$ 1 mG in both Models 1 and 2.
The average field strength also exceeds the value expected only from the shock compression.
Does the growth of average field strength come only from the strong amplification at the narrow transition layers?
Our answer is no, since the filling factor of the region where the magnetic field is amplified to the level of 1 mG is small.
Furthermore, as seen in Fig.3 and 4, the magnetic fields are amplified to tens or hundreds micro-Gauss broadly in the downstream.
This growth of the average field strength may be called "turbulent dynamo" (Batchelor 1950; Cho \& Vishniac 2000; Cho \& Lazarian 2003) that leads the magnetic field amplification through the stretching of field line by turbulent eddies (vortexes).
Of course the origin of the eddy is the rippling of the shock front that is due to the density inhomogeneity of the multi-phase preshock gas.
We discuss more details about the turbulent dynamo in the next section by taking spectra of the turbulent postshock medium.

The difference between the average field strengths between Models 1 and 2 is larger than the difference between the maximum field strengths.
This is simply due to the difference in initial field amplification by the simple shock compression.
The level of average field strengths is roughly an order of magnitude smaller than the maximum.
Therefore, in some regions where the magnetic field is amplified to 1 mG, about a half of preshock kinetic energy is converted to magnetic energy, but on average the conversion factor is only 1\% or less depending on the angle of the global shock normal and the magnetic field. 

\begin{figure}[t]
\epsscale{1.}
\plotone{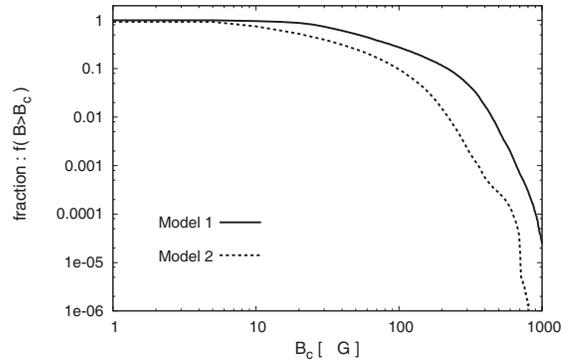}
\caption{
Cumulative volume fraction of the region where magnetic field strength 
 is larger than the threshold $B_{\rm C}$ in the shell 
 at the times depicted in Fig. \ref{f3}.
}
\label{f6}
\end{figure}

In Fig. \ref{f6} we plot the volume fractions of the regions where magnetic field strength is larger than $B$ in the shell at the times depicted in Fig. \ref{f3} and Fig. \ref{f4}.
The volume fractions where $|B|\ge 100\,\,\mu$G and $500\,\,\mu$G are respectively 27.6\% and 0.5\% in Model 1 and 9.6\% and 0.03\% in Model 2.

\subsubsection{Spatial Profiles and Power spectra}\label{spct}
\begin{figure}
\epsscale{1.}
\plotone{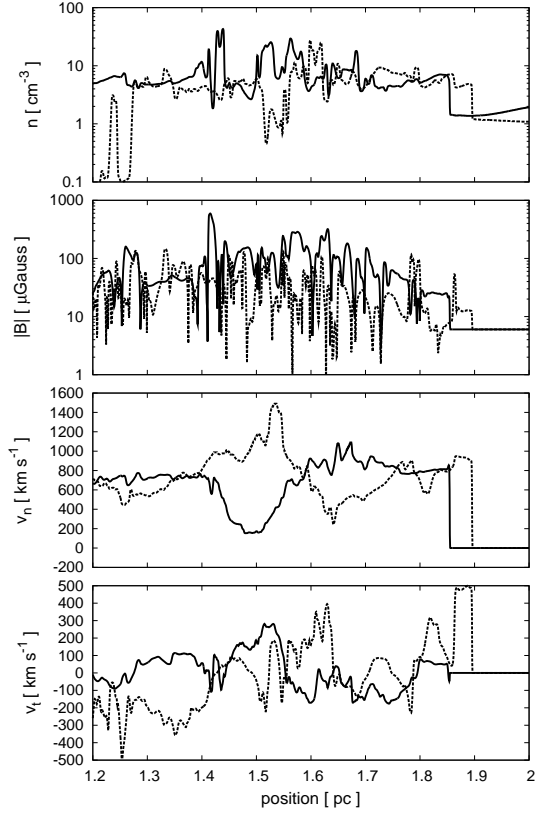}
\caption{
Cross section profiles along $y$-axis at $x=1.0$ pc of Model 1 ({\it solid}) and $x$-axis at $y=1.0$ pc of Model 2 ({\it dotted}).
The number density, magnetic field strength, velocity component normal to the shock ($v_{y}$ for Model 1 and $v_{x}$ for Model 2), and velocity component tangential to the shock ($v_{x}$ for Model 1 and $v_{y}$ for Model 2) are plotted from the top to bottom panel.
The shock front is located around position 1.85 pc (Model 1) and 1.90 pc (Model 2).
The left and the right sides of these are downstream and upstream, respectively.
}
\label{f7}
\end{figure}

In Fig. \ref{f7} we show cross section profiles of Model 1 along the $y$-axis at $x=1.0$ pc ({\it solid}) and of Model 2 along the $x$-axis at $y=1.0$ pc ({\it dotted}).
The number density, magnetic field strength, velocity component normal to the shock ($v_{y}$ for Model 1 and $v_{x}$ for Model 2), and velocity component tangential to the shock ($v_{x}$ for Model 1 and $v_{y}$ for Model 2) are plotted from the top to bottom panel.
The shock front is located around the position $y=1.85$ pc (Model 1) and $x=1.90$ pc (Model 2).
The left and right sides of the shock front are downstream and upstream, respectively.
From the density cross-section, we can see many clumps in the shell with well defined boundaries.
Most of the clumps are the shocked CNM and some of the density jumps are generated by secondary shocks in the turbulence.
The velocity cross sections also show large amplitude fluctuations, in which discontinuous jumps due to the secondary shocks can be seen.

\begin{figure}[t]
\epsscale{1.}
\plotone{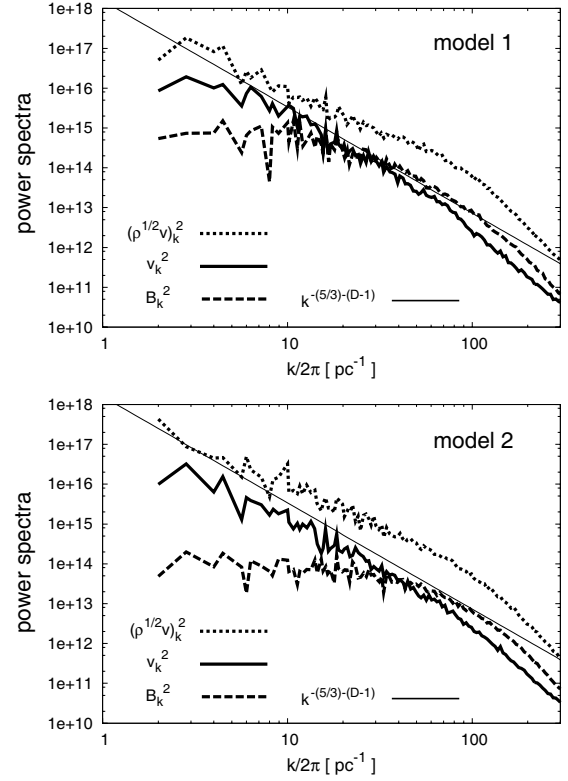}
\caption{
Power spectra of the velocity $v_k^2$ ({\it thick solid}), magnetic field $B_k^2/4\pi$ ({\it dashed}), and $(\sqrt{\rho}\,v)_k^2$ ({\it dotted}).
The top panel shows the spectra in the region $x\in [0.75,1.25]$ and $y\in [1.3,1.8]$ of Model 1 at $t=1425$ yr, and the bottom panel shows those in the region $x\in [1.3,1.8]$ and $y\in [0.75,1.25]$ of Model 2 at $t=1508$ yr.
The thin solid lines show Kolmogorov spectra $v_k^2\propto k^{-(5/3)-(D-1)}$ at $D=2$.
The ratio of $B_k^2/4\pi$ and $(\sqrt{\rho}v)_k^2$ indicates the ratio of the magnetic and kinetic energies at scale $k$.
}
\label{f8}
\end{figure}

\begin{figure}[t]
\epsscale{1.}
\plotone{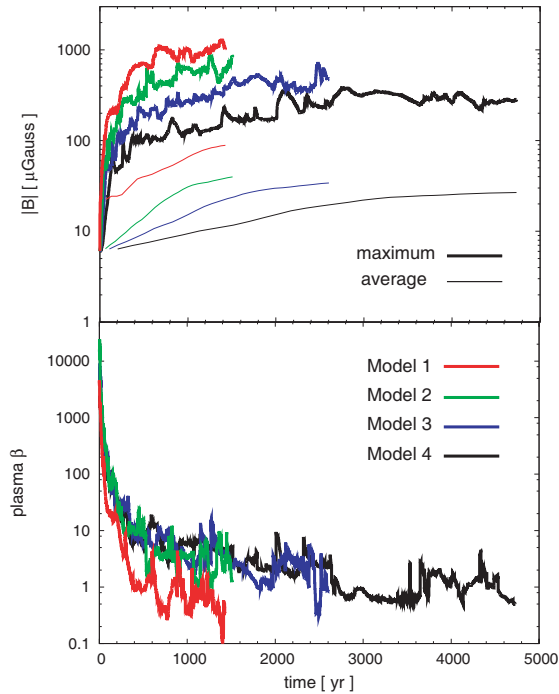}
\caption{
Evolutions of the maximum ({\it thick lines}) and average ({\it thin lines}) field strengths ({\it top}), and the local plasma $\beta$s at the point where the magnetic field strength is maximum ({\it bottom}).
The red, green, blue, and black lines are the results of Models 1, 2, 3, and 4, respectively.
}
\label{f9}
\end{figure}

In order to understand the statistical properties of the turbulent fluctuations in the shocked shell, it is helpful to observe their power spectra. 
Fig. \ref{f8} shows the power spectra of the velocity $v_k^2$, magnetic field $B_k^2/4\pi$, and $(\sqrt{\rho}v)_k^2$ obtained from the data of Model 1 in the region $x\in [0.75,1.25]$ and $y\in [1.3,1.8]$ ({\it top}), and Model 2 in the region $x\in [1.3,1.8]$ and $y\in [0.75,1.25]$ ({\it bottom}).
Note that the shell integrals in $k$ space of $B_k^2/8\pi$, and $(\sqrt{\rho}v)_k^2/2$ give, respectively, the magnetic energy and the kinetic energy.
Thus the ratio of $B_k^2/4\pi$ and $(\sqrt{\rho}v)_k^2$ indicates the ratio of the magnetic and kinetic energies at scale $k$.

The spectra of two models show qualitatively similar profiles.
This is due to the fact that the turbulence in the shells is super-Alfv\'enic in both cases.
In the case of the super-Alfv\'enic turbulence, as mentioned in the previous section, vortexes (eddies) cause stretching and amplify magnetic field not only at the transition layers but also in diffuse region (turbulent dynamo).
In the following we compare the spectra taking from our results and that of other simulations in which the driven turbulent dynamo is studied.

In the large scales ($k/2\pi\lesssim 100$ pc$^{-1}$), the velocity power spectra show the Kolmogorov spectrum $v_{k}^2\propto k^{-(5/3)-(D-1)}$ ($D=3$ in three-dimensional case and $D=2$ in two-dimensional case), and the spectra of magnetic fields are flatter than those of velocities.
These properties are also indicated from the simulations of the simple, one-phase, super-Alfv\'enic driven turbulence (see, e.g. Cho \& Lazarian 2003).
The coincidence of the large scale characteristics of $v_{k}^2$ and $B_{k}^2$ between our two-phase results and other one-phase results is not surprising, since the character of two-phase medium arises only on small scales owing to the smallness of the CNM clumps.
The only difference between the two models on large scales is that the power spectrum of the magnetic field in Model 1 has larger amplitude than that of Model 2.
This can be also seen in the bottom panels of Fig. \ref{f3} and \ref{f4}, i.e., the magnetic field in Model 1 has larger structures than in Model 2.
In Model 1, the timescale of the interaction between the shock wave and CNM clumps is longer than that of Model 2 owing to the filamentary structure of the CNM clumps that would cause the larger eddies and thus larger structures of the magnetic field.
In contrast to the velocity, the power spectra of $\sqrt{\rho}v$ are slightly flatter than the Kolmogorov spectrum.
This is due to the small scale structures of the shocked CNM clumps, since the existence of the delta-function-like structures of the density makes the power spectra flatter than the Kolmogorov spectrum (e.g., Hennebelle \& Audit 2007).

On small scales ($k/2\pi\gtrsim 100$ pc$^{-1}$), the velocity spectra decline more steeply than Kolmogorov.
The level of the magnetic power $B_k^2$ and the kinetic power $(\sqrt{\rho}v)_k^2$ are closer than on the large scale.
In the case of  the transonic, super-Alfv\'enic turbulence (Cho \& Lazarian 2003) and the incompressible, super-Alfv\'enic turbulence (Kida et al. 1991: Cho \& Vishniac 2000), the power of the magnetic field is reported to be larger than that of the velocity field on small scales.
Note that in the case of one-phase turbulence the ratio of the velocity and magnetic field power spectra gives the ratio of the kinetic and magnetic energy spectra, since the density is normalized to unity. 
In our case, however, the magnetic field power spectrum does not dominate the kinetic power spectrum even on small scales, although the difference is small.
A result similar to ours was also reported in Padoan \& Nordlund (1999) in which highly supersonic, super-Alfv\'enic turbulence  was studied.
Thus, whether or not the magnetic power dominates the kinetic power on small scales seems to depend on the property of dynamics. 

So far, we have not discussed time evolutions of the spectra.
In the case of our simulation, it is difficult to analyze the detailed time dependence of the power spectra, since shocked shell is too thin to take the spectra at early stage.
In the case of the results of Cho \& Vishniac (2000) and Cho \& Lazarian (2003), until the growth of average magnetic field saturates, the power spectrum of magnetic field shows inverse cascade, i.e., the scale of magnetic field becomes larger as the field becomes stronger.
In our case, the average field strength does not show clear saturation, though the growth rate is small compared to the earlier stage (see the top panel of Fig. 9).
Thus, the average field strength has some possibility to grow further and the scale at which the spectra change their properties ($k \sim 100$ at the moment taking the spectra in Fig. 8) possibly becomes larger.
Thus, it remains imperative to do long term calculations for studying the time evolution of the turbulence spectra.

\subsubsection{Dependence on Shock Strength}\label{ps}

\begin{figure}[t]
\epsscale{1.}
\plotone{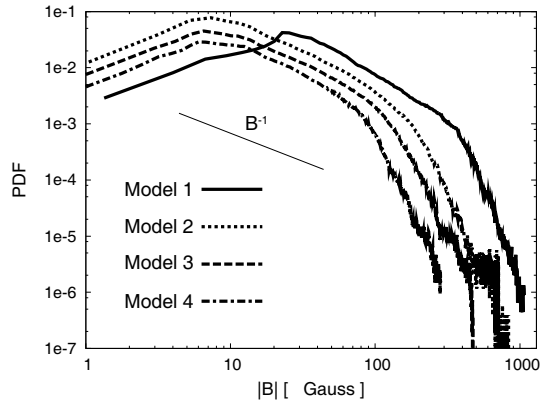}
\caption{
Probability distribution functions (PDF) of the magnetic field strength in the shocked shell.
The solid, dotted, dashed, and dot-dashed lines are the results of Models 1 (at $t=1425$), 2 (at $t=1425$), 3 (at $t=2602$), and 4 (at $t=4731$), respectively.
}
\label{f10}
\end{figure}

In Figure \ref{f9}, we show the results of the parameter study.
The top panel represents the evolutions of the maximum ({\it thick lines}) and average ({\it thin lines}) field strengths, and the bottom panel represents the local plasma $\beta$s at the point where the magnetic field strength is maximum.
The red, green, blue, and black lines are the results of Models 1, 2, 3, and 4 ,respectively.
The top panel shows that the magnetic field strengths decrease with increasing thermal pressure of the hot plasma or decreasing strength of the shock waves.
As indicated in \S \ref{s31}, in the case of Models 1 and 2, the velocity dispersion of the turbulence is on the order of sound speed of the post shock medium.
In the case of Models 3 and 4, the velocity dispersions in the shocked shell are respectively 240 km s$^{-1}$ and 130 km s$^{-1}$, which are on the order of the sound speeds in each shocked shell (see equation [\ref{ss}]).
If one assumes that, as discussed in \S \ref{s32}, the maximum magnetic field strength is determined by the strength of the turbulent flow that is comparable to the sound speed, the local plasma $\beta$ where magnetic field strength is maximum should be on the order of unity.
The bottom panel clearly verifies this expectation.

The velocity and magnetic field power spectra do not show any qualitative difference between Models 1 and 2.
The probability distribution functions (PDF) of the magnetic field strength in the shocked shell (Fig. \ref{f10}) indicate that there exists a range in which the PDFs show the power law dependence of the field strength ($P[B]\propto B^{-p}$) with the index $p\sim1$.
This power law range is extended to higher field strength with increasing shock speed and increasing angle of the pre-shock magnetic field direction to the normal of the plane of the global shock wave. 

\section{Summary and Discussions}
We have examined the MHD simulations of the shocked two-phase medium in two-dimensional geometry.
We have shown that the shocked shell becomes turbulent owing to the preshock density inhomogeneity and the magnetic field amplification taking place in the shell.
This is consistent with the work of Giacalone \& Jokipii (2007).
In the cases of Models 1 and 2, in which the propagation speeds of the shocks correspond to the Sedov-Taylor stage of supernova blast wave, the magnetic field strength is amplified upto the order of 1mG that is roughly determined by the condition $\beta\sim1$.

The preshock density inhomogeneity in our model originates in the thermal instability that is naturally and ubiquitously expected in the ISM (Inoue \& Inutsuka 2008).
Thus, the process of the magnetic field amplification investigated in this paper is also expected in real SNRs. 
Recent X-ray observations of SNRs by Uchiyama et al. (2007) and Uchiyama \& Aharonian (2008) discovered synchrotron X-ray hot spots or filaments.
The magnetic fields are amplified up to $\sim1$ mG there, if the flux-decreasing timescale of the X-ray hot spots, which is on the order of a few years, is determined by the synchrotron cooling timescale of accelerated electrons, $t_{\rm synch}\simeq 1.5\,(B/\rm{mG})^{-1.5}\,(\epsilon/\rm{keV})^{-0/5}$ year, where $\epsilon$ is the energy of emitted photon.
In the following, we compare our simulation results with the observed characterstics of the X-ray hot spots.

(1) {\it Spatial scale:}
The typical scale length of the observed X-ray hot spots is $\ell\sim0.05$ pc, which is comparable to the scale of the regions where the magnetic field strength is amplified to the order of 1 mG in our simulation (see, Fig. \ref{f5}, where regions depicted in red have field strengths larger than 0.6 mG).
This scale length is determined by the width of the transition layers between the CNM clumps and surrounding WNM ($\sim0.1$--0.01 pc) at which the strongest shear is induced and amplifies the magnetic field. 

(2) {\it Location of the X-ray hot spot:}
The observed X-ray hot spots seem to be located at more than 0.1 pc behind the shock front, although the projection effect should be taken into account.
This fact can be naturally explained in our scenario (see bottom panels of Fig. \ref{f3} and Fig. \ref{f4} and Fig.\ref{f7}).
However, it may be difficult for high-energy electrons, which accelerated at the shock front and emitting synchrotron X-rays, to move from the shock front to the emitting region (hot spot) because of the synchrotron cooling.
Moreover, the crossing time of the hot spots for such electrons, $\ell/v$ where $v$ is the flow velocity at the hot spot, is much longer than the observed variability timescale of a few years.
These facts imply that the electrons responsible for the X-ray hot spots are not accelerated at the shock front, but accelerated {\it in situ} or at least very nearby compared to the shock front.
A possible acceleration site is the secondary shocks arising from the turbulent flow (e.g., collisions of turbulent flows) in the shocked shell.
As indicated in \S \ref{spct}, our simulations show the generation of many secondary shocks in the shocked shell (see, Fig. \ref{f7}). 

(3) {\it Variability timescale and electron acceleration:}
The maximum magnetic field strength in our simulation is large enough to explain the flux decreasing timescale of a few years.
However, it is still uncertain whether the secondary shock that sweeps the region with strong magnetic field can accelerate electrons enough to make the hot spot.
In order to confirm this, further studies are necessary taking into account the acceleration of electrons at the secondary shocks.

The filling factor of the regions with high-magnetic field (B~1mG) that correspond to the hot spots is less than 1\% in our results.
However, we emphasize here that it depends on the number of the CNM clumps embedded in the diffuse WNM.
If we consider the region of ISM including much larger number of CNM clumps, a shock wave that piles up such ISM to form SNR is expected to show more hot spots.
In particular, huge amounts of HI CNM clumps are expected around molecular clouds (see, e.g. Blitz et al. 2006 for observation and Koyama \& Inutsuka 2002; Inoue \& Inutsuka 2008; Hennebelle et al. 2008 and Heitsch et al. 2008 
for simulations of molecular cloud formation).
For example, SNR RXJ1713 is suggested to be interacting with molecular clouds (Fukui et al. 2008), and thus, may show many hot spots than our simulations.

The evidence for magnetic field amplification in SNR has been also obtained from the thickness of X-ray filaments that shows typically $B\sim100$ microGauss (Bamba et al. 2003, 2005a, 2005b; Vink \& Laming 2003).  
Since the X-ray filaments have apparently coherent features on parsec scales, it seems to be refrecting the "average" field strength in the SNR.
The average field strength in our simulation shows tens to a hundred micro-Gauss (see, Fig. 9) that possibly explains the X-ray filaments.
Note that the thickness $\sim 0.01$ pc of the X-ray filaments would be indicating the spatial distribution of accelerated electrons.  
At this moment this is not described in our simulations omitting nonthermal effects.
Obviously the nonthermal effects (e.g., Lucek \& Bell 2000) should be also important in understanding the origin of the X-ray filaments.

Our simulations predict the spatial power spectrum of magnetic field distribution in the SNR shell that is slightly shallower than the Kolmogorov spectrum on the scales $k< 100$ pc$^{-1}$ and slightly steeper than the Kolmogorov on the scales $k> 100$ pc $^{-1}$ as is the case of the super Alfv\'enic turbulence (see, Fig. \ref{f8}).
Such fluctuations of the magnetic field should be reflected as a spatial inhomogeneity of the synchrotron radiations as the synchrotron X-ray images of SNRs depicted in Uchiyama et al. (2007) and Uchiyama \& Aharonian (2008).
Although the inhomogeneity of the radiation reflects the inhomogeneous distribution of accelerated particles, it may not be easy to directly measure the spatial power spectrum of the magnetic field from observations.

Magnetic field strength on the order of mG is also indicated in the center of the Galaxy where spectacular filamentary structures are observed with polarized radio emission.  
Although the mechanism of creation of large-scale filamentary structure with non-thermal emission remains to be found, our result provides a hint for the origin of $\sim$ mG field strength because the Galactic center is expected to be occupied by many supernova remnants, and hence, should be the site for the magnetic field amplification studied in this paper.
Spreading out of the locally amplified magnetic field into large scale volume is also expected in starburst galaxies that inevitably experience concurrent supernova explosions. 
Thus, it might be interesting to study the long-term evolution of the strongly magnetized turbulent interstellar medium with the effects of multiple strong shock waves in such environments. 

In this paper we have performed simulations in two-dimensional geometry in order to keep the spatial resolution. 
In principle, it is imperative to perform similar high resolution simulations in three dimensional geometry to analyze the realistic three-dimensional phenomena.
We expect, however, that the possible difference is not significant, since the statistical properties of the turbulence in our simulation can be understood consistently with the three-dimensional simulations of super Alfv\'enic turbulence (see, section \ref{spct}).
We also neglect the effect of explicit magnetic diffusivity in the calculation of stage 2, since the resistivity of the collisionless plasma in supernova remnants is highly unknown.
It is known that the strength of magnetic field amplification by turbulent dynamo depends on the resistivity or magnetic Reynolds number $R_{\rm m}$, which determine the efficiency of magnetic field dissipation mainly via 
reconnection, i.e., larger resistivity (smaller Rm) leads weaker amplification of magnetic field.
However we expect that the saturation level of magnetic field strength does not change, in the cases where $R_{\rm m}$ is roughly lager than 100 (see, Fig. 2 of Cho \& Vishniac 2000).
In our simulation the reconnection of magnetic field takes place at the scale of numerical grid.
The magnetic Reynolds number $R_{\rm m}$ can be estimated to the order of the cell number in the numerical domain that is $\sim$ 1,000 in our simulation.  
Therefore, we think that our results should not change unless the resistivity in supernova remunants is huge to leads $R_{\rm m} < 100$.

\acknowledgments
T.I. thanks T. K. Suzuki for the helpful discussion of MHD turbulence and numerical scheme.
Numerical computations were carried out on NEC SX-9 and XT4 at Center for Computational Astrophysics (CfCA) of National Astronomical Observatory of Japan, and NEC SX-8 at YITP in Kyoto University.

This work is supported in part by a Grant-in-aid from the Ministry of Education, Culture, Sports, Science, and Technology (MEXT) of Japan, No.18740153 and No.19047004 (R.Y), and No.15740118, No.16077202, and No.18540238 (S.I.).

\end{document}